\documentclass[aps,twocolumn,showpacs,preprintnumbers,superscriptaddress,amsmath,amssymb,prl]{revtex4}
\usepackage{color}
\usepackage{natbib}
\usepackage{bm, graphicx, amsmath}
\usepackage{hyperref}
\usepackage{braket}
\graphicspath{{./images/}}

\begin{document}

\title{Quantum retrodiction made fully symmetric}
\author{Dov Fields}
\affiliation{Department of Physics and Astronomy, Hunter College and the Graduate Center of the City University of New York, 10065 New York, NY, USA}
\author{Abdelali Sajia}
\affiliation{The Center for Quantum Science and Engineering, Stevens Institute of Technology, 07030 Hoboken, NJ, USA}
\author{J\'anos A. Bergou}
\affiliation{Department of Physics and Astronomy, Hunter College and the Graduate Center of the City University of New York, 10065 New York, NY, USA}
\begin{abstract}
	Quantum retrodiction is a time-symmetric approach to quantum mechanics with applications in a number of important problems. One of the major challenges to its more widespread applicability is the restriction of its symmetric formalism to unbiased sources. The main result of this paper is to develop a general theory yielding a symmetric formalism for arbitrary sources.  We then highlight on a specific example, by presenting the optimal solution to the retrodiction problem that is dual to unambiguous state discrimination, how the generalized approach works. We also show how this formalism leads to a symmetric formulation of the communication channel between Alice and Bob and point to the intrinsic connection between retrodiction and the no-signaling principle.
\end{abstract}
\date{\today}
\maketitle
Quantum retrodiction, a time symmetric approach to quantum mechanics, was initially proposed by Aharonov and collaborators in a series of papers \cite{Aharonov1964,Aharonov1984,Aharonov1991} and discussed by Penfield \cite{Penfield1966}.  It was, however, the seminal work by Barnett \emph{et al.,} that put the method on solid foundation, applying Bayesian inference to relate forward (predictive) and backward (retrodictive) conditional probabilities \cite{Barnett1999}. Many of the subsequent works, clarifying additional aspects of the theory, are summarized in Ref. \cite{Barnett2014}.  After the foundational groundworks, quantum retrodiction has found widespread applications in areas including quantum optics \cite{Pegg1999,Barnett2000,Amri11}, atomic systems \cite{Barnett2000b,Jeffers2002},  open systems \cite{Barnett2001a,Pegg2002}, communication schemes \cite{Hillery2016}, and characterization of measurement devices \cite{Pegg2002b,Amri2011,Dauria2011}. Recent research in this area includes experimental implementations \cite{Yuan2018}, quantum imaging \cite{Speirits2017}, quantum state tomography \cite{Scerri2019}, and entropic relations for retrodictive measurements \cite{Budini2018}. 

A key quantity in all of these works is the \emph{source function}. Assume that, in a standard quantum mechanical scenario, an observer has the knowledge that the system under investigation is in state $\rho_{i}$ ($i=1,2,\hdots$) with the prior probabilities $\eta_{i}$, such that $\sum_{i} \eta_{i} = 1$.  From the observer's perspective, the state of the system can best be described by the source function, 
\begin{equation}
\Omega = \sum_{i} \eta_{i} \rho_{i} \ .
\label{source}
\end{equation}
The source is \emph{unbiased} if it satisfies the condition $\sum_{i} \eta_{i}\rho_{i} = \frac{1}{D}I$, where $D$ is the dimension and $I$ is the identity operator for the Hilbert space of the system, meaning that the source is in a maximally mixed (unbiased) state. Otherwise, the source is biased. 

It has been noticed in \cite{Barnett1999} and \cite{Pegg2002} that, for the case of an \emph{unbiased source}, a fully time-symmetric and operational theory of retrodiction can be developed. The major limitation for more widespread applications of this operational approach is, of course, the restriction to unbiased sources. The main goal of this letter is to remove this restriction and to extend the theory to systems with general (biased) sources. After a brief overview of the method in \cite{Barnett1999} and \cite{Pegg2002}, we present our main result. We introduce a transformation, uniquely defined by the source function, that makes the theory fully symmetric for general sources. With this symmetric formulation, retrodiction can now serve as the dual to completely general predictive problems. In the second part of the letter, we demonstrate the power of the method on the \emph{dual} (retrodictive) problem to unambiguous state discrimination (UD). Curiously, unlike the UD problem, its dual problem is not a state identification but a state exclusion problem. We want to emphasize that this problem could not even be addressed using the approach restricted to unbiased sources.  In the last part of the letter we highlight the inherent connection between retrodiction and the no-signaling principle and prove that the communication channel based on unambiguous discrimination is fully symmetric. Ultimately, by introducing the fully time-symmetric formulation of quantum retrodiction, we extend the theory past its previous limitations and provide a powerful tool for analyzing quantum communication channels and quantum measurements.

We begin by considering, in detail, the two-party communication scheme described before Eq. \eqref{source}. Alice prepares systems in one of the states from the set $a = \left\{a_{i}= \rho_{i} |i=1,2,\hdots\right\}$ with the prior probabilities $\eta_{i}$ and sends one system at a time over to Bob. The set of states and their priors are also known to Bob.  When Bob receives a system, he performs a measurement on it, using the set of detectors $b = \left\{b_{j} = \Pi_{j} |j=1,2,\hdots\right\}$, in order to get information about its state. The detectors, in general, form a POVM (Positive Operator Valued Measure, projective measurement being a special case), with the properties $\sum_{j} \Pi_{j} = I$ and $\Pi_{j} \geq 0$, where $I$ is again the identity operator. Recalling Born's rule, Alice can predict the conditional probability that Bob's measurement outcomes is $j$ (a click in detector $j$ is recorded),
\begin{equation}
	P\left( b_{j}|a_{i} \right) = Tr\left( \Pi_{j}\rho_{i} \right) \ ,
	\label{predictiveprobability}
\end{equation}
given that state $i$ was prepared. This is the conventional predictive use of standard quantum theory. 
 
We can also view the measurement from Bob's perspective which is the basis for the retrodictive approach. From this perspective, there is a set of detectors $\left\{ \Pi_{j} \right\}$ that click with probabilities 
\begin{equation}
P(b_{j}) \equiv \mu_{j} = \sum_{i}Tr\left( \Pi_{j}\rho_{i} \right)\eta_{i} \equiv Tr\left( \Pi_{j}\Omega \right) ,
\label{mretdef}
\end{equation}
for the set of states, $\{\rho_{i}\}$, sent by Alice.  From his measurement data, Bob can \emph{infer} the conditional probability, $P\left( b_{j}|a_{i} \right)$, that Alice sent a particular state, given that detector $j$ clicked. $P\left( b_{j}|a_{i} \right)$ is the retrodictive probability. Employing Bayes' theorem, $P\left( a_{i}|b_{j} \right) P\left( b_{j} \right) = P\left( b_{j}|a_{i} \right)P\left( a_{i} \right)$, immediately yields

\begin{eqnarray}
	P\left( a_{i}|b_{j} \right) &=& \frac{P\left( b_{j}|a_{i} \right)P\left( a_{i} \right)}{P\left( b_{j} \right)} \nonumber \\
	&=& \frac{Tr\left( \Pi_{j}\rho_{i} \right)\eta_{i}}{\sum_{i}Tr\left( \Pi_{j}\rho_{i} \right)\eta_{i}} .
	\label{retrodictiveprobabiliy}
\end{eqnarray}

As stated in the introduction, it was noticed in \cite{Barnett1999,Pegg2002} that for an \emph{unbiased source}, $\Omega = \sum_{i} \eta_{i}\rho_{i} = \frac{1}{D}I$, the theory can be cast to a form with full symmetry between the prediction and retrodiction. By introducing $\rho_{j}^{ret} \equiv \frac{\Pi_{j}}{Tr\left( \Pi_{j} \right)}$ and $\eta_{j}^{ret} \equiv Tr(\Pi_{j})/D = \mu_{j}$, it is easy to show that the source function for retrodiction is also unbiased, $\sum_{j} \eta_{j}^{ret}\rho_{j}^{ret} = \frac{1}{D}I$ holds. Further, by introducing $\Pi_{i}^{ret} \equiv D\eta_{i}\rho_{i}$,  $\sum_{j} \Pi_{j} ^{ret}= I$ and $\Pi_{j} ^{ret}\geq 0$ also hold, defining a POVM for retrodiction. There is full symmetry between the predictive sets, $a^{p} = \left\{ \rho_{i} \right\}$ and $b^{p} = \left\{ \Pi_{j} \right\}$, and the retrodictive sets, $b^{r} = \left\{ \rho_{i}^{ret} \right\}$ and $a^{r}=\left\{ \Pi_{j}^{ret} \right\}$. The retrodictive probability can now be written as

\begin{equation}
	P\left( a_{i}|b_{j} \right) = Tr\left( \Pi_{i}^{ret}\rho_{j}^{ret} \right) ,
	\label{retrodictivesym}
\end{equation}
which is just Born's rule for the retrodictive operators. There is now also full symmetry between the predictive and retrodictive probabilities. Both now have operational meaning. Thus, the theory proposed in \cite{Barnett1999}, provides a solid foundation for treating the predictive and retrodictive approaches on equal footing. 

However, as mentioned before, there is one major limitation of this, otherwise very elegant and powerful, theory. It is restricted to unbiased sources. We now show that this restriction can be eliminated and develop a completely general theory of retrodiction, as our main result.

To this end, we define the transformation,
\begin{eqnarray}
	\Pi_{i}^{ret} &\equiv& \Omega^{-\frac{1}{2}}\eta_{i}\rho_{i}\Omega^{-\frac{1}{2}} \label{Pretdef} , \\
	\rho_{j}^{ret} &\equiv& \frac{\sqrt{\Omega}\Pi_{j}\sqrt{\Omega}}{\mu_{j}} .
	\label{rhoretdef}
\end{eqnarray}
with the help of the source function, $\Omega$, that maps the states $\left\{ \rho_{i} \right\}$ to a set of positive operators $\left\{ \Pi_{i}^{ret} \right\}$ that have the detector normalization and the detectors $\left\{\Pi_{j} \right\}$ to a set of positive operators $\left\{ \rho_{j}^{ret} \right\}$ that have the state normalization, exactly as their predictive counterparts, 
\begin{equation}
\sum_{i} \Pi_{i}^{ret} = I \ \ \  \mbox{and} \ \ \ Tr\left( \rho_{j}^{ret} \right) = 1 .
\label{Piret}
\end{equation} 
In terms of these operators, Eq. \eqref{retrodictivesym} now holds for the retrodictive probability for \emph{arbitrary} sources, exhibiting complete symmetry with the predictive probability, Eq. \eqref{predictiveprobability}. Further, the predictive and retrodictive source states are identical,
\begin{equation}
	\sum_{i}\eta_{i}\rho_{i} = \Omega = \sum_{j}\mu_{j}\rho_{j}^{ret} ,
	\label{retconst}
\end{equation}
but otherwise arbitrary, \emph{not restricted to multiples of the identity}. Finally, we note that, in the special case when $\Omega = \frac{1}{D}I$, the general definitions, Eqs. \eqref{Pretdef} and \eqref{rhoretdef}, reduce to the ones given in \cite{Barnett1999}. 

The transformations described in Eqs. \eqref{Pretdef} and \eqref{rhoretdef}, with the properties given in Eqs. \eqref{Piret} and \eqref{retconst}, constitute our main result. Together, they ensure that Eq. \eqref{retrodictivesym} is now operationally defined for arbitrary sources. Prediction and retrodiction are fully symmetric, operational, and we arrive at a general time-symmetric formulation of quantum mechanics.  

\begin{figure}[ht]
	\centerline{
	\setlength{\unitlength}{1pt}
	\begin{picture}(180,200)(0,0)
		\put(0,0){\includegraphics[height = 200pt,width=180pt]{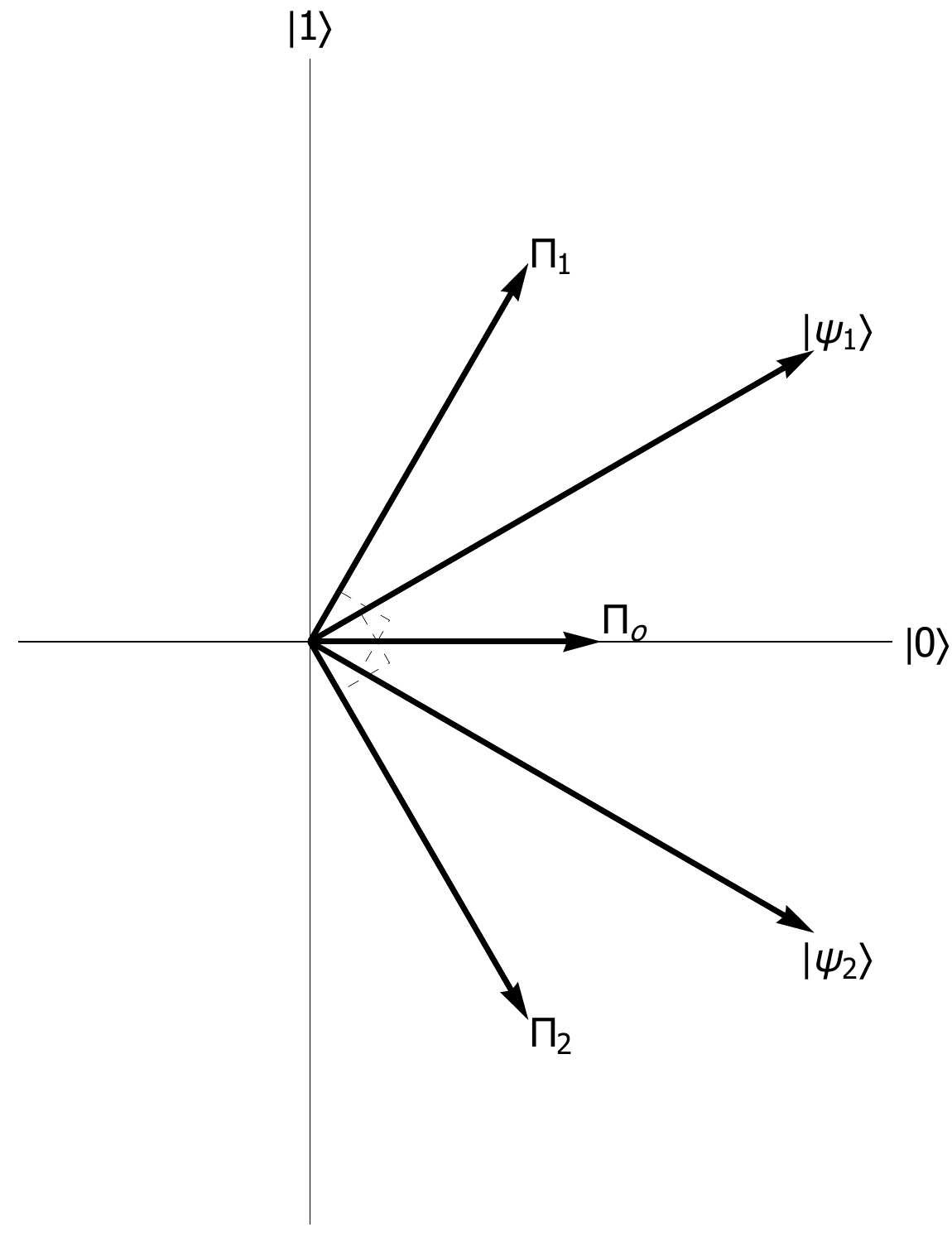}}
	\end{picture}}
	\caption{Graphical representation of the optimal Unambiguous Discrimination setup for equal priors using the  predictive approach of standard quantum mechanics. Real states correspond to vectors in a 2-dimensional plane. Bob's detector $\Pi_{1}$ ($\Pi_{2}$) is orthogonal to the input state $\ket{\psi_{2}}_{b}$ ($\ket{\psi_{1}}_{b}$). The inconclusive detector $\Pi_{0}$ is directed symmetrically between the two input states. The figure also shows Alice's measurement, preparing the input states for Bob. She performs a standard projective measurement along the $\ket{0}_{a}$ and $\ket{1}_{a}$ directions in the frame of her qubit, on the state in Eq. \eqref{ABshared}. When she finds $\ket{0}_{a}$ ($\ket{1}_{a}$), Bob 's qubit will be in the $\ket{\psi_{1}}_{b}$ ($\ket{\psi_{2}}_{b}$) state. There is no symmetry between the roles of Alice and Bob.}
	\label{fig:preimg}
\end{figure}

Next, we demonstrate the power of the generalized theory on the example of the dual (retrodictive) problem to unambiguous state discrimination (UD), \cite{Ivanovic1987,Dieks1988,Peres1987,Jaeger1995}. (See \cite{Barnett2009} and \cite{Bergou2010}, for recent reviews on state discrimination.) In the standard UD problem, Alice prepares a qubit in one of two known pure states, $\left\{ \ket{\psi_{1}}, \ket{\psi_{2}} \right\}$ with prior probabilities $\left\{ \eta_{1}, \eta_{2} \right\}$, and sends it to Bob. Bob's task is to discriminate between these states using a POVM, $\Pi = \left\{ \Pi_{1},\Pi_{2}, \Pi_{0} \right\}$ ($\sum_{i}\Pi_{i} = I$ and $\Pi_{i} \geq 0$), satisfying 
\begin{equation}
\braket{\psi_{1}|\Pi_{2}|\psi_{1}} = \braket{\psi_{2}|\Pi_{1}|\psi_{2}} = 0, 
\label{UDcondition}
\end{equation}
 i.e., requiring that the $\Pi_{i}$ detector clicks only if the state $\ket{\psi_{i}}$ is sent ($i=1,2$). This will ensure that, when detector $\Pi_{1}$ or $\Pi_{2}$ clicks, the measurement is unambiguous. The price to pay is that Bob has to allow for an inconclusive (or failure) outcome of his measurement, $\Pi_{0}$, that can occur for either input. The goal is to maximize the average success probability,
\begin{equation}
	P_{s} = \eta_{1}Tr\left( \Pi_{1}\ket{\psi_{1}}\bra{\psi_{1}} \right) + \eta_{2}Tr\left( \Pi_{2}\ket{\psi_{2}}\bra{\psi_{2}} \right) ,
	\label{probsuccessUD}
\end{equation}
which also minimizes the average probability of failure. The optimal solution is well known \cite{Jaeger1995} (the optimal set-up will be shown later, in Fig. \ref{fig:preimg}). 

For the dual problem, using the retrodictive formalism, the average success probability of retrodiction is given in terms of the transformed operators, Eqs. \eqref{Pretdef} and \eqref{rhoretdef}, as
\begin{eqnarray}
	P_{s} &=& \mu_{1}Tr\left( \Pi_{1}^{ret}\rho_{1}^{ret} \right) + \mu_{2}Tr\left( \Pi_{2}^{ret}\rho_{2}^{ret} \right).	
	\label{retoptcondition}
\end{eqnarray}
In the predictive UD problem, one is given the set of states and the goal is to find the measurement that will optimize the average success probability of prediction, \eqref{probsuccessUD}. In the dual problem, one is given the measurement operators and the goal is to find the states that will optimize the average success probability of retrodiction, \eqref{retoptcondition}.

To derive the optimal success probability, we assume, with no loss of generality, that we are working in a 2-dimensional Hilbert space. Introducing the states
\begin{equation}
\ket{\phi_{i}^{ret}} = \Omega^{-\frac{1}{2}}\sqrt{\eta_{i}}\ket{\psi_{i}} , \ \ \ \ i=1,2 ,
\label{retbasis}
\end{equation}
we can write the pure state version of Eq. \eqref{Pretdef} as $\Pi_{i}^{ret} = \ket{\phi_{i}^{ret}}\bra{\phi_{i}^{ret}}$. Inserting this in the normalization of the retrodictive detectors, Eq. \eqref{Piret}, gives $\ket{\phi_{1}^{ret}}\bra{\phi_{1}^{ret}} +  \ket{\phi_{2}^{ret}}\bra{\phi_{2}^{ret}} = I$. It then follows that the states in \eqref{retbasis} are orthonormal (for a full proof see the Supp. Mat., \cite{suppmat}), so they form what we call the retrodictive basis. 

Next, we make use of the UD condition, \eqref{UDcondition}. Combining it with Eqs. \eqref{mretdef}--\eqref{retrodictivesym}, yields $\eta_{2}Tr\left( \Pi_{1}\ket{\psi_{2}}\bra{\psi_{2}} \right) = \mu_{1}Tr\left( \rho_{1}^{ret}\Pi_{2}^{ret} \right) = 0$, or $Tr\left( \rho_{1}^{ret}\Pi_{2}^{ret} \right) = Tr\left( \rho_{2}^{ret}\Pi_{1}^{ret} \right) = 0$. Using this and $\Pi_{1}^{ret} + \Pi_{2}^{ret} =I$, gives 
\begin{equation}
Tr\left( \rho_{1}^{ret}\Pi_{1}^{ret} \right) = Tr\left( \rho_{2}^{ret}\Pi_{2}^{ret} \right) = 1 .
\label{retoptUD}
\end{equation}
This implies $\rho_{j}^{ret} = \ket{\phi_{j}^{ret}}\bra{\phi_{j}^{ret}}$ (\cite{suppmat}). So, from \eqref{retconst}, 
\begin{equation}
	\mu_{1}\ket{\phi_{1}^{ret}}\bra{\phi_{1}^{ret}} + \mu_{2}\ket{\phi_{2}^{ret}}\bra{\phi_{2}^{ret}} + \mu_{0}\rho_{0}^{ret} = \Omega .
	\label{retconstraint}
\end{equation}
Using \eqref{retbasis} gives the matrix elements of $\Omega$ in the retrodictive basis, $\Omega_{ij} = \sqrt{\eta_{i}\eta_{j}} \braket{\psi_{i}|\psi_{j}}$, $i=1,2$.  Thus, with $|\braket{\psi_{1}|\psi_{2}}| = \cos(\theta)$, the matrix form of $\Omega$ becomes
\begin{equation}
	\Omega =
	\begin{pmatrix}
		\eta_{1} & \sqrt{\eta_{1}\eta_{2}}\cos(\theta) \\ \sqrt{\eta_{1}\eta_{2}}\cos(\theta) & \eta_{2} 
	\end{pmatrix} .
	\label{Omegamatrix}
\end{equation}
Inserting \eqref{Omegamatrix} into \eqref{retconstraint}, and rearranging, we obtain
\begin{equation}
	\mu_{0}\rho_{0}^{ret} =
	\begin{pmatrix}
		\eta_{1} - \mu_{1} & \sqrt{\eta_{1}\eta_{2}}\cos\left( \theta \right) \\ \sqrt{\eta_{1}\eta_{2}}\cos\left( \theta \right) & \eta_{2} - \mu_{2}
	\end{pmatrix} ,
	\label{altretconstraint}
\end{equation} 
We note that the same expression appears in a paper by Barnett and Andersson \cite{Barnett2001}, as pointed out after Eq. \eqref{ameasuredreduced}, deriving optimal UD from the no-signaling condition. (We will discuss the inherent connection between retrodiction and no-signaling later.) 

Combining Eqs. \eqref{retoptcondition} and \eqref{retoptUD}, we can now state the dual problem as follows. For a given set of states $\left\{\ket{\phi_{1}^{ret}},\ket{\phi_{2}^{ret}}\right\}$ and a source $\Omega$, optimize the function,
\begin{equation}
	P_{s} = \mu_{1} + \mu_{2} ,
	\label{retsuccess}
\end{equation}
subject to the constraint $\mu_{0}\rho_{0}^{ret} \geq 0$. Using the constraint, it is straightforward to obtain the optimal success probability (for details of the derivation see \cite{suppmat}),
\begin{equation}
	P_{s} =
	\begin{cases}
				1 - 2\sqrt{\eta_{1}\eta_{2}} s & \frac{1}{2} < \eta_{max} < \frac{1}{1 + s^{2}} \\
		\eta_{max}\left( 1 - s^{2} \right) &\frac{1}{1+ s^{2}} \leq \eta_{max}
	\end{cases}.
	\label{retoptsol}
\end{equation}
Here $\eta_{max(min)}$ is the larger (smaller) of $\eta_{1},\eta_{2}$ and $s \equiv \cos(\theta)$ (to save notation). Finally, for the optimal solution, $det\left( \rho_{0}^{ret} \right) = 0$, i.e., $\rho_{0}^{ret}$ is a pure state, $\rho_{0}^{ret} = \ket{\phi_{0}^{ret}}\bra{\phi_{0}^{ret}}$, where $\ket{\phi_{0}^{ret}} = \frac{1}{\sqrt{2}}\left( \ket{\phi_{1}^{ret}} + \ket{\phi_{2}^{ret}} \right)$.

We want to highlight two lessons from this example. First, the problem in this example could not even be addressed, much less solved, using the method of \cite{Barnett1999,Pegg2002} because the source function is general, not unbiased. The theory is now completely general: It can address the dual problem to any standard predictive problem. Second, it is also interesting to note that the dual problem to UD is not an UD problem. Rather, it is a kind of a state exclusion problem \cite{Bandyopadhyay2014}, where a given measurement outcome excludes one or more of the possible states. In our case, a click in the retrodictive detector $\Pi_{1(2)}^{ret}$ excludes the retrodictive state $\rho_{2(1)}^{ret}$. The twist is that the optimization is not over the measurement but in finding the optimum decomposition of the source function that maximizes the average probability of success for a given set of detectors. This is a very satisfactory picture: UD is about the optimal identification of the states that make up the source, its dual problem is about the optimal exclusion of the states that make up the source.

After demonstrating how the general theory works, we now return to the remark made after Eq. \eqref{altretconstraint}. In order to understand how retrodiction and no-signaling are related, we will slightly reformulate the predictive UD problem. In the standard version, Alice sends Bob one of two states, $\ket{\psi_{1}}$ or $\ket{\psi_{2}}$, with probabilities $\eta_{1},\eta_{2}$, respectively. This can also be accomplished if, instead, Alice and Bob initially share the entangled state, 
\begin{equation}
\ket{\Psi}_{ab} = \sqrt{\eta_{1}}\ket{0}_{a}\ket{\psi_{1}}_{b} + \sqrt{\eta_{2}}\ket{1}_{a}\ket{\psi_{2}}_{b} .
\label{ABshared}
\end{equation}
If Alice measures her qubit in the $\left\{\ket{0}_{a},\ket{1}_{a}\right\}$ basis and finds $\ket{0}_{a}$ ($\ket{1}_{a}$), then Bob's qubit will be in the state $\ket{\psi_{1}}_{b}$ ($\ket{\psi_{2}}_{b}$), with probability $\eta_{1}$ ($\eta_{2}$). The optimal set-up for Alice's state preparing and Bob's state discriminating measurement is depicted in Fig. \ref{fig:preimg}.

The no-signaling condition requires that if Bob performs an optimal UD measurement to distinguish between $\ket{\psi_{1}}_{b},\ket{\psi_{2}}_{b}$, before Alice performs her own measurement, she should be able to gain no information about the outcome of Bob's measurement. Before Bob's measurement, the state of the qubit in Alice's possession can be described by the reduced density matrix, $\rho_{a} = Tr_{b}\left( \ket{\Psi}_{ab}\bra{\Psi} \right)$ in the $\{\ket{0}_{a},\ket{1}\}_{a}$ basis. From \eqref{ABshared}, it is easy to see that the matrix form of $\rho_{a}$ is identical to the matrix form of $\Omega$ in \eqref{Omegamatrix} in the retrodictive basis. The reason why these two matrices must be identical will become clear from the discussion surrounding Eq. \eqref{symcommunication}.

If Bob's measurement succeeds, then Alice's state is the pure state, $\ket{0}_{a}$ or $\ket{1}_{a}$, corresponding to Bob's measurement outcome. This outcome happens with Bob's respective success probabilities, $p_{1}, p_{2}$, where $p_{i}$ is the probability that Bob's measurement succeeds, given the state $\ket{\psi_{i}}_{b}$. If Bob's measurement fails, Alice's state can be described by some unknown mixed state $\rho_{0}$, which is the outcome with probability $p_{0}$. Given this, if Alice does not know the outcome of Bob's measurement, she can describe her state as,
\begin{equation}
	\tilde{\rho}_{a} = p_{1}\ket{0}{_{a}\bra{0}} + p_{2}\ket{1}{_{a}\bra{1}} + p_{0}\rho_{0}.
	\label{ameasuredreduced}
\end{equation}
The no-signaling condition requires that Alice gain no information from Bob's measurement, or $\rho_{a} = \tilde{\rho}_{a}$, giving the earlier constraint, Eq. \eqref{altretconstraint}, when we make the substitution $p_{i} \rightarrow \mu_{i}$. 
\par
\begin{figure}[ht]
	\centerline{
	\setlength{\unitlength}{1pt}
	\begin{picture}(160,200)(0,0)
		\put(0,0){\includegraphics[height = 200pt, width = 160pt]{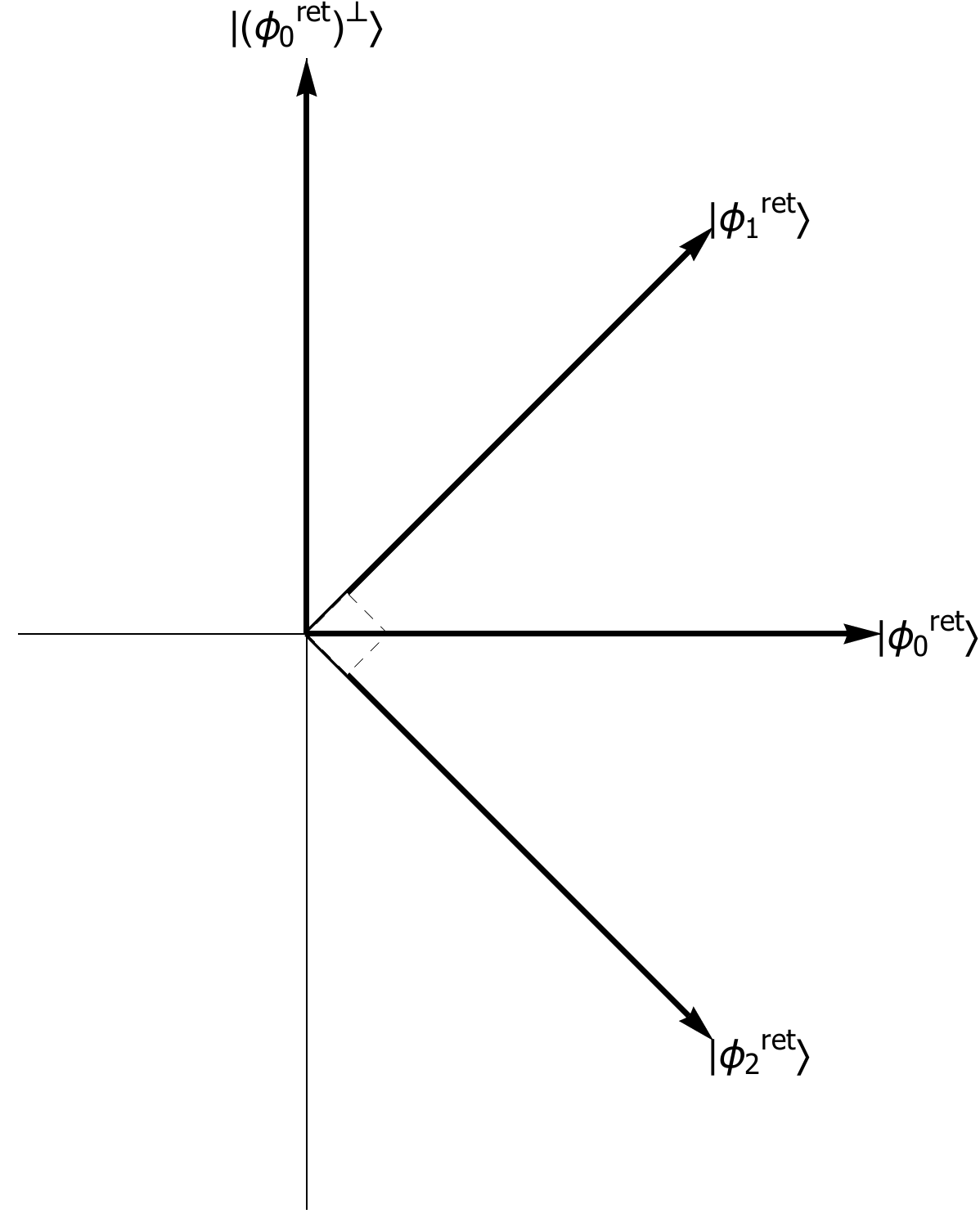}}
	\end{picture}}
	\caption{ Graphical representation of the symmetric communication channel, based on the retrodictive model. Real states correspond to vectors in a 2-dimensional plane. For state preparation, Alice performs standard projective measurements on $\ket{\Psi}_{ab}$, first line in \eqref{symcommunication}, in her retrodictive basis, $\ket{\phi_{1}^{ret}}_{a}$ and $\ket{\phi_{2}^{ret}}_{a}$. After Alice's state preparation measurement, Bob performs the state discrimination measurement, shown in Fig. \ref{fig:preimg}. Alternatively, Bob can perform the state preparation measurement first on $\ket{\Psi}_{ab}$, second line in \eqref{symcommunication}, using his retrodictive basis, $\ket{\phi_{1}^{ret}}_{b}$ and $\ket{\phi_{2}^{ret}}_{b}$. After Bob's state preparation measurement, now Alice performs the state discrimination measurement, shown in Fig. \ref{fig:preimg}. So, either party can be the sender or the receiver. There is \emph{full symmetry} between the roles of Alice and Bob.}
	\label{fig:retimg}
\end{figure}

In order to understand why the no-signaling constraint is identical to the retrodictive constraint, we first note that the relevant feature in Eq. \eqref{ABshared} is that Alice entangles two orthogonal states of her qubit with two non-orthogonal states of Bob's qubit. When she performs a measurement in the basis defined by her orthogonal states, it will prepare states for Bob to discriminate. We also note that Alice can use any orthonormal basis (ONB) for state preparation, her reduced density matrix will always be the same, \eqref{Omegamatrix}, \emph{in the basis she uses for state preparation}. In particular, this will also hold if she uses her retrodictive basis for state preparation. In this basis only, the following expression holds 
\begin{eqnarray}
	\ket{\Psi}_{ab}^{sym} &=& \sqrt{\eta_{1}}\ket{\phi_{1}^{ret}}_{a}\ket{\psi_{1}}_{b} + \sqrt{\eta_{2}}\ket{\phi_{2}^{ret}}_{a}\ket{\psi_{2}}_{b} \nonumber \\
	&=& \sqrt{\eta_{1}}\ket{\psi_{1}}_{a}\ket{\phi_{1}^{ret}}_{b} + \sqrt{\eta_{2}}\ket{\psi_{2}}_{a}\ket{\phi_{2}^{ret}}_{b}.
	\label{symcommunication}
\end{eqnarray}
The surprising feature is the equality of the two lines, i.e., in the retrodictive basis, the communication channel between Alice and Bob is fully symmetric under the $A \leftrightarrow B$ swap \cite{suppmat}. Furthermore, taking the trace over Alice's qubit, using the first line, gives that Bob's reduced state is equal to the source function, $\rho_{b}=\Omega$, as it should. The crucial point is that taking the trace over Bob's qubit, Alice's reduced state is also equal to the source function, $\rho_{a}=\Omega$. So, finally, we conclude that, since the matrix form of Alice's reduced state is the same in the basis she used for state preparation, it must be the matrix of $\Omega$ in the retrodictive basis, which is precisely what Eq. \eqref{Omegamatrix} is.Hence, the constraint on retrodiction is equivalent to the no-signaling condition. The set-up for the symmetric communication channel, based on the retrodictive approach, is depicted in Fig. \ref{fig:retimg}.

\emph{Summary} In this paper we extended the theory of Quantum Retrodiction to completely general systems and eliminated the restriction to unbiased sources. Most importantly, the theory is fully symmetric between prediction and retrodiction. Previous versions of the theory of Quantum Retrodiction, restricted to unbiased sources and without the full symmetry, have already found a large number of applications, so our work should allow Quantum Retrodiction to be applicable to an even broader, in fact, unrestricted range of problems. One avenue to explore is to use it to set up dual problems to standard predictive optimization problems. As a first step in this direction, we presented a derivation of the optimal solution for the dual (retrodictive) problem to unambiguous discrimination. In general, there is the potential that the theory could assist in optimization problems that currently have no known solution. Finally, the demonstration of the inherent connection between our general theory and the no-signaling principle which, in turn, establishes that the communication channel between Alice and Bob can be made symmetric for \emph{any} source, is a major step towards elevating the no-signaling condition, and hence the present fully time-symmetric retrodictive approach, from a working principle to a fundamental postulate of quantum theory.

\emph{Acknowledgement and disclaimer} Research was sponsored by the Army Research Laboratory and was accomplished under
Cooperative Agreement Number {\bf{W911NF-20-2-0097}}. The views and conclusions contained in this document are
those of the authors and should not be interpreted as representing the official policies, either expressed or implied, of
the Army Research Laboratory, or the U.S. Government. The U.S. Government is authorized to reproduce and
distribute reprints for Government purposes notwithstanding any copyright notation herein.

\end{document}